\documentclass{aa}
\usepackage{txfonts} \usepackage{graphicx}
\usepackage{color}

\begin{document}

\title{Nonradial and radial period changes in the $\delta$ Scuti star 4 CVn\\
I. 700+ nights of photometry}
 
\author{M.~Breger\inst{1,2}}

\institute{Institut f\"ur Astrophysik der Universit\"at Wien, T\"urkenschanzstr. 17, A--1180 Wien, Austria\\
        \email{michel.breger@univie.ac.at}
\and
Department of Astronomy, University of Texas, Austin, TX 78712, USA}

\date{Received date; accepted date}

\abstract 
{The nature of period and amplitude changes in nonradial pulsators is presently unknown.} {It is therefore important to examine the correlations between these changes in stars with a large number of simultaneously excited pulsation modes. However, the small amplitudes require extensive high-precision photometry covering many years.} {We present 702 nights of high-precision photometry of the evolved $\delta$
Sct variable 4~CVn obtained from 2005 - 2012 with a dedicated telescope.} {We detected 64 frequencies, of which 38 can be identified as combinations and harmonics. The relative amplitudes of the combination frequencies are similar to those found in 44 Tau and show no evidence for resonant mode coupling. Significant period and amplitude changes are detected for the dominant modes. The known prograde and retrograde modes show period changes with opposite signs, while the radial mode
exhibits only small, cyclical period changes. For each mode, the period changes are constant over the eight years and range from (1/P)dP/dt = -16 x 10$^{-6}$ to 13 x 10$^{-6}$.
On the other hand, the amplitude variations show no systematic behavior between different pulsation modes.}{The behavior of the prograde, axisymmetric, and retrograde modes indicate a constant decrease in the rotational splitting over the eight years.}
 
\keywords{Stars: variables. $\delta$~Sct -- Stars: oscillations -- Stars: individual: 4~CVn -- Techniques: photometric}

\titlerunning{Nonradial and radial period changes in 4 CVn I. 700+ nights of photometry}

\maketitle

\section{Introduction}

The  $\delta$ Scuti stars are short-period pulsators with spectral types A and F mostly in the main-sequence or immediate post-main-sequence stage of evolution. These stars pulsate with radial and nonradial pressure and gravity modes. High-amplitude radial pulsation generally requires slow rotation in the high-amplitude $\delta$ Scuti stars (known as HADS), while the vast majority of the pulsators possess dominant low-amplitude nonradial pulsation modes and medium to fast rotation.

Almost all studied nonradially pulsating $\delta$ Scuti stars show amplitude and  period variations with typical timescales of years and decades. Some of these variations are not true changes intrinsic to the star, but are observational in nature. Two such examples are light-time effects in binary stars (Murphy et al. 2014) and  beating between two close frequencies resulting in beat periods less than a year (Breger \& Bischof 2002).

The nature and physical origins of the common variations with longer timescales of years and decades are not known at this time. However, the variety and size of the period variations exclude an evolutionary origin (Breger \& Pamyatnykh 1998).

The recent high-precision photometric measurements obtained from spacecraft such as $\it{KEPLER}$ have yielded additional information on period and amplitude variability, for example, the amplitudes of
many $\delta$ Scuti stars show rotational modulation due to stellar spots. Furthermore, the multiyear $\it{KEPLER}$ photometry of KIC 8054146 has revealed that the amplitude and period variations of a few parent modes are responsible for similar changes in a large number of other (child) modes due to resonant mode coupling (Breger \& Montgomery 2014). Nevertheless, even for this star, the reason for the changes in the parent modes is still unknown.

The available analyses  indicate that the study of the properties and physical origin of the period and amplitude variations require extensive observations covering many years or decades.
One of the most promising stars for such an examination is the evolved $\delta$ Sct variable 4~CVn (HR~4715 = HD~107904 = AI CVn, F3III-IV).  Jones \& Haslam (1966) discovered the variability
of this star. This led to a large number of photometric studies, of which some were multisite campaigns to minimize the effects of aliasing. The publications prior to 2000 are summarized in Table 1 of Breger (2000).
Since then, additional photometry from 1991 was made available by R. J. Dukes (Breger, Davis, \& Dukes 2008). Schmid et al. (2014) analyzed the line-profile variations of 4 CVn from over 2000 high-dispersion spectra obtained from 2008 to 2011 and determined the prograde or retrograde nature of seven pulsation modes. Schmid et al. also discovered that 4 CVn is an eccentric binary system with an orbital period of 124.44 $\pm$ 0.03 d. These mode identifications are important to look for systematic behavior between different types of pulsation modes excited in the same star.

\begin{table}[h]
\caption{New photometry of 4~CVn}
\label{table1}
\begin{center}
\begin{tabular}{lcc}
\noalign{\smallskip}    
\hline\hline
\noalign{\smallskip}
Year & Number of nights & Date (HJD)\\
& & 245 0000+ \\
\noalign{\smallskip}
\hline
\noalign{\smallskip}
2005    &       62      &       3371.92  -      3560.67 \\
2006    &       63      &       3781.79  -      3922.70 \\
2007    &       86      &       4084.94  -      4278.72 \\
2008    &       87      &       4449.97  -      4638.74 \\
2009    &       76      &       4800.98  -      5008.73 \\
2010    &       96      &       5163.01  -      5382.65 \\
2011    &       108     &       5546.95  -      5738.69 \\
2012    &       124     &       5906.94  -      6099.74 \\
\noalign{\smallskip}
\hline
\end{tabular}
\end{center}
\end{table}

\begin{table*}
\begin{center}
\caption{Multifrequency solution of 4\,CVn}
\begin{tabular}{lrrlrrlrrlrr}                                                                                                                                                                                   
\noalign{\smallskip}                                                                                                                                                                                                                                                                                            \hline                                                                                                                                                                                  
\hline                                                                                                                                                                                  
\noalign{\smallskip}                                                                                                                                                                                    
\multicolumn{3}{c}{Frequency} & ID & \multicolumn{2}{c}{Average amplitudes}&\multicolumn{3}{c}{Frequency} & ID & \multicolumn{2}{c}{Average amplitudes}\\                                                                 &&&&$v$ filter & $y$ filter& & &  & & $v$ filter & $y$ filter\\
& cd$^{-1}$ &$\mu$Hz&  &millimag&millimag & & cd$^{-1}$ &$\mu$Hz&  &millimag&millimag\\                                                                                                                                                                                         
\noalign{\smallskip}                                                                                                                                                                                    
\hline                                                                                                                                                                                  
\noalign{\smallskip}                                                                                                                                                                                    
$f_{1}$ &       8.594   &       99.47   &               &       25.4    &       17.3    &       \hspace{15mm}$f_{33}$   &       1.928   &       22.32   &       $f_{6}$-$f_{2}$ &       1.0     &       0.6     \\
$f_{2}$ &       5.048   &       58.43   &               &       25.7    &       16.7    &       \hspace{15mm}$f_{34}$   &       2.745   &       31.77   &       $f_{1}$-$f_{3}$ &       1.0     &       0.7     \\
$f_{3}$ &       5.850   &       67.71   &               &       18.6    &       12.8    &       \hspace{15mm}$f_{35}$   &       12.025  &       139.17  &       $f_{2}$+$f_{6}$ &       1.0     &       0.6     \\
$f_{4}$ &       5.532   &       64.02   &               &       14.9    &       9.9     &       \hspace{15mm}$f_{36}$   &       1.185   &       13.72   &       $f_{5}$-$f_{10}$        &       1.1     &       0.5     \\
$f_{5}$ &       7.376   &       85.37   &               &       10.9    &       7.2     &       \hspace{15mm}$f_{37}$   &       1.069   &       12.37   &       $f_{9}$-$f_{2}$ &       1.0     &       0.6     \\
$f_{6}$ &       6.976   &       80.75   &               &       9.9     &       6.4     &       \hspace{15mm}$f_{38}$   &       15.970  &       184.84  &       $f_{1}$+$f_{5}$ &       0.8     &       0.8     \\
$f_{7}$ &       7.552   &       87.40   &               &       8.8     &       6.0     &       \hspace{15mm}$f_{39}$   &       10.580  &       122.45  &       $f_{2}$+$f_{4}$ &       1.0     &       0.6     \\
$f_{8}$ &       6.680   &       77.32   &               &       8.6     &       5.9     &       \hspace{15mm}$f_{40}$   &       10.898  &       126.13  &       $f_{2}$+$f_{3}$ &       0.9     &       0.6     \\
$f_{9}$ &       6.117   &       70.80   &               &       8.8     &       5.7     &       \hspace{15mm}$f_{41}$   &       10.281  &       118.99  &       $f_{4}$+$f_{12}$        &       0.9     &       0.6     \\
$f_{10}$        &       6.191   &       71.65   &               &       7.1     &       4.5     &       \hspace{15mm}$f_{42}$   &       1.618   &       18.72   &       $f_{1}$-$f_{6}$ &       0.9     &       0.5     \\
$f_{11}$        &       6.440   &       74.54   &               &       6.8     &       4.5     &       \hspace{15mm}$f_{43}$   &       16.032  &       185.55  &       $f_{5}$+$f_{25}$        &       0.8     &       0.6     \\
$f_{12}$        &       4.749   &       54.97   &               &       6.8     &       4.4     &       \hspace{15mm}$f_{44}$   &       13.343  &       154.44  &       $f_{1}$+$f_{12}$        &       0.8     &       0.5     \\
$f_{13}$        &       5.136   &       59.45   &               &       2.1     &       1.3     &       \hspace{15mm}$f_{45}$   &       14.711  &       170.27  &       $f_{1}$+$f_{9}$ &       0.8     &       0.5     \\
$f_{14}$        &       6.905   &       79.92   &               &       1.7     &       1.1     &       \hspace{15mm}$f_{46}$   &       12.600  &       145.83  &       $f_{2}$+$f_{7}$ &       0.8     &       0.5     \\
$f_{15}$        &       6.403   &       74.11   &               &       1.7     &       1.0     &       \hspace{15mm}$f_{47}$   &       10.096  &       116.85  &       2$f_{2}$        &       0.7     &       0.5     \\
$f_{16}$        &       6.749   &       78.12   &               &       1.4     &       1.0     &       \hspace{15mm}$f_{48}$   &       15.274  &       176.79  &       $f_{1}$+$f_{8}$ &       0.7     &       0.5     \\
$f_{17}$        &       5.314   &       61.51   &               &       1.5     &       0.9     &       \hspace{15mm}$f_{49}$   &       11.723  &       135.68  &       $f_{4}$+$f_{10}$        &       0.7     &       0.5     \\
$f_{18}$        &       6.108   &       70.69   &               &       1.4     &       0.9     &       \hspace{15mm}$f_{50}$   &       11.649  &       134.82  &       $f_{4}$+$f_{9}$ &       0.7     &       0.5     \\
$f_{19}$        &       4.499   &       52.07   &               &       1.4     &       0.8     &       \hspace{15mm}$f_{51}$   &       15.571  &       180.22  &       $f_{1}$+$f_{6}$ &       0.7     &       0.5     \\
$f_{20}$        &       7.873   &       91.12   &               &       1.1     &       0.7     &       \hspace{15mm}$f_{52}$   &       13.779  &       159.48  &       $f_{5}$+$f_{15}$        &       0.7     &       0.4     \\
$f_{21}$        &       4.712   &       54.53   &               &       1.1     &       0.7     &       \hspace{15mm}$f_{53}$   &       1.368   &       15.83   &       $f_{9}$-$f_{12}$        &       0.7     &       0.4     \\
$f_{22}$        &       5.184   &       60.01   &               &       1.1     &       0.6     &       \hspace{15mm}$f_{54}$   &       10.922  &       126.42  &       $f_{1}$-$f_{2}$+$f_{5}$ &       0.6     &       0.4     \\
$f_{23}$        &       5.572   &       64.49   &               &       0.9     &       0.6     &       \hspace{15mm}$f_{55}$   &       14.353  &       166.12  &       $f_{6}$+$f_{5}$ &       0.6     &       0.4     \\
$f_{24}$        &       0.468   &       5.41    &               &       1.0     &       0.5     &       \hspace{15mm}$f_{56}$   &       15.249  &       176.49  &       $f_{5}$+$f_{20}$        &       0.5     &       0.4     \\
$f_{25}$        &       8.655   &       100.18  &               &       0.7     &       0.5     &       \hspace{15mm}$f_{57}$   &       0.400   &       4.63    &       $f_{5}$-$f_{6}$ &       0.6     &       0.3     \\
$f_{26}$        &       1.699   &       19.66   &               &       0.6     &       0.4     &       \hspace{15mm}$f_{58}$   &       13.953  &       161.49  &       2$f_{6}$        &       0.5     &       0.4     \\
\multicolumn{6}{l}{Combinations and harmonics}                          &       \hspace{15mm}$f_{59}$   &       16.146  &       186.87  &       $f_{1}$+$f_{7}$ &       0.4     &       0.2     \\
$f_{27}$        &       13.642  &       157.90  &       $f_{1}$+$f_{2}$ &       1.7     &       1.1     &       \hspace{15mm}$f_{60}$   &       17.188  &       198.94  &       2$f_{1}$        &       0.3     &       0.2     \\
$f_{28}$        &       2.328   &       26.95   &       $f_{5}$-$f_{2}$ &       1.6     &       0.9     &       \hspace{15mm}$f_{61}$   &       22.947  &       265.59  &       $f_{1}$+$f_{6}$+$f_{5}$ &       0.3     &       0.2     \\
$f_{29}$        &       13.226  &       153.08  &       $f_{3}$+$f_{5}$ &       1.1     &       0.7     &       \hspace{15mm}$f_{62}$   &       23.522  &       272.25  &       $f_{1}$+$f_{5}$+$f_{7}$ &       0.2     &       0.2     \\
$f_{30}$        &       1.143   &       13.23   &       $f_{10}$-$f_{2}$        &       1.0     &       0.7     &       \hspace{15mm}$f_{63}$   &       24.565  &       284.31  &       2$f_{1}$+$f_{5}$        &       0.2     &       0.2     \\
$f_{31}$        &       1.632   &       18.89   &       $f_{8}$-$f_{2}$ &       1.1     &       0.7     &       \hspace{15mm}$f_{64}$   &       20.619  &       238.64  &       $f_{1}$+$f_{2}$+$f_{6}$ &       0.2     &       0.1     \\
$f_{32}$        &       11.451  &       132.54  &       $f_{2}$+$f_{15}$        &       1.1     &       0.7     \\                                                                                                                                                                              
\noalign{\smallskip}
\hline
\end{tabular}
\end{center}
\end{table*}

\section{New photometry of 4~CVn}

\begin{figure*}
\includegraphics[bb=10 100 750 550,width=\textwidth,clip]{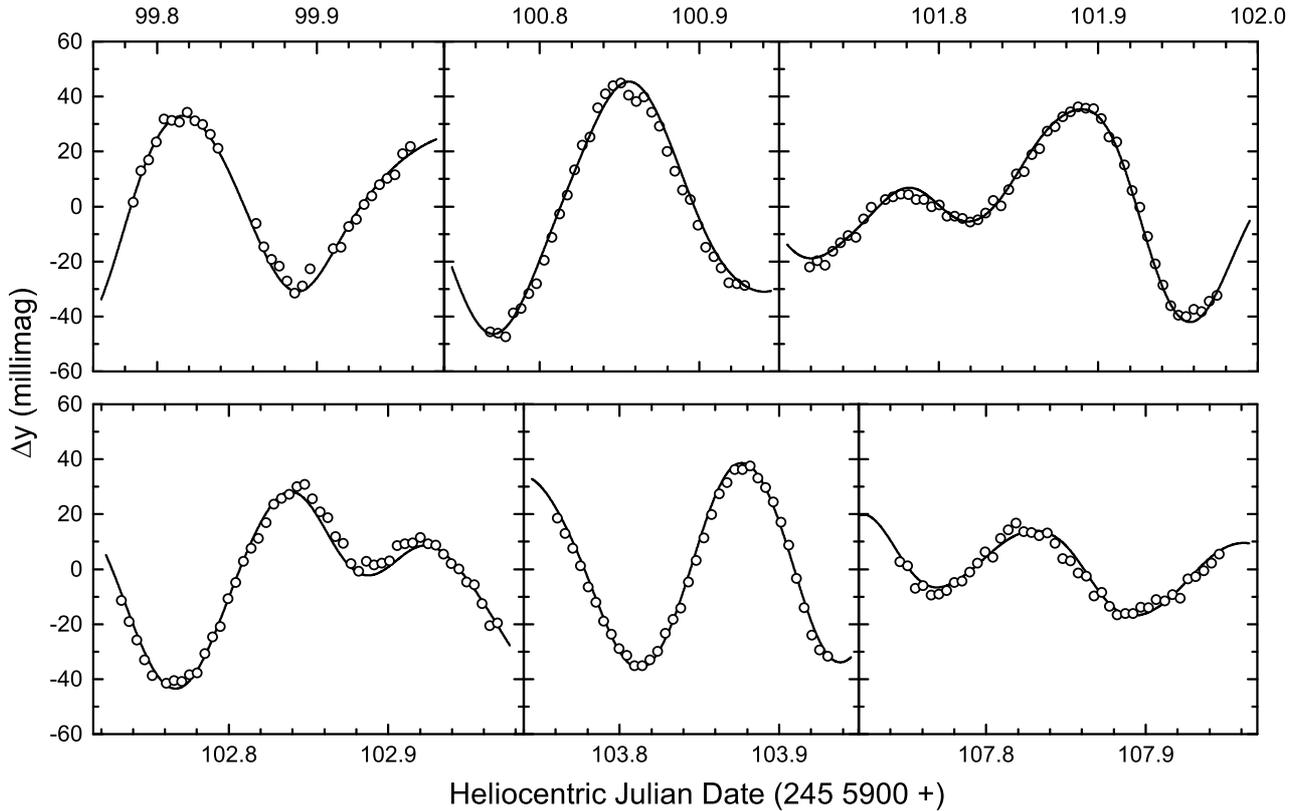}
\caption{Variability of 4 CVn for 6 typical nights in the middle of the 2012 campaign, which covered 124 nights. The curves shown are the 64-frequency fits obtained from all of the available data.}

\end{figure*}

\begin{figure*}
\includegraphics[bb=30 30 810 510,width=\textwidth,clip]{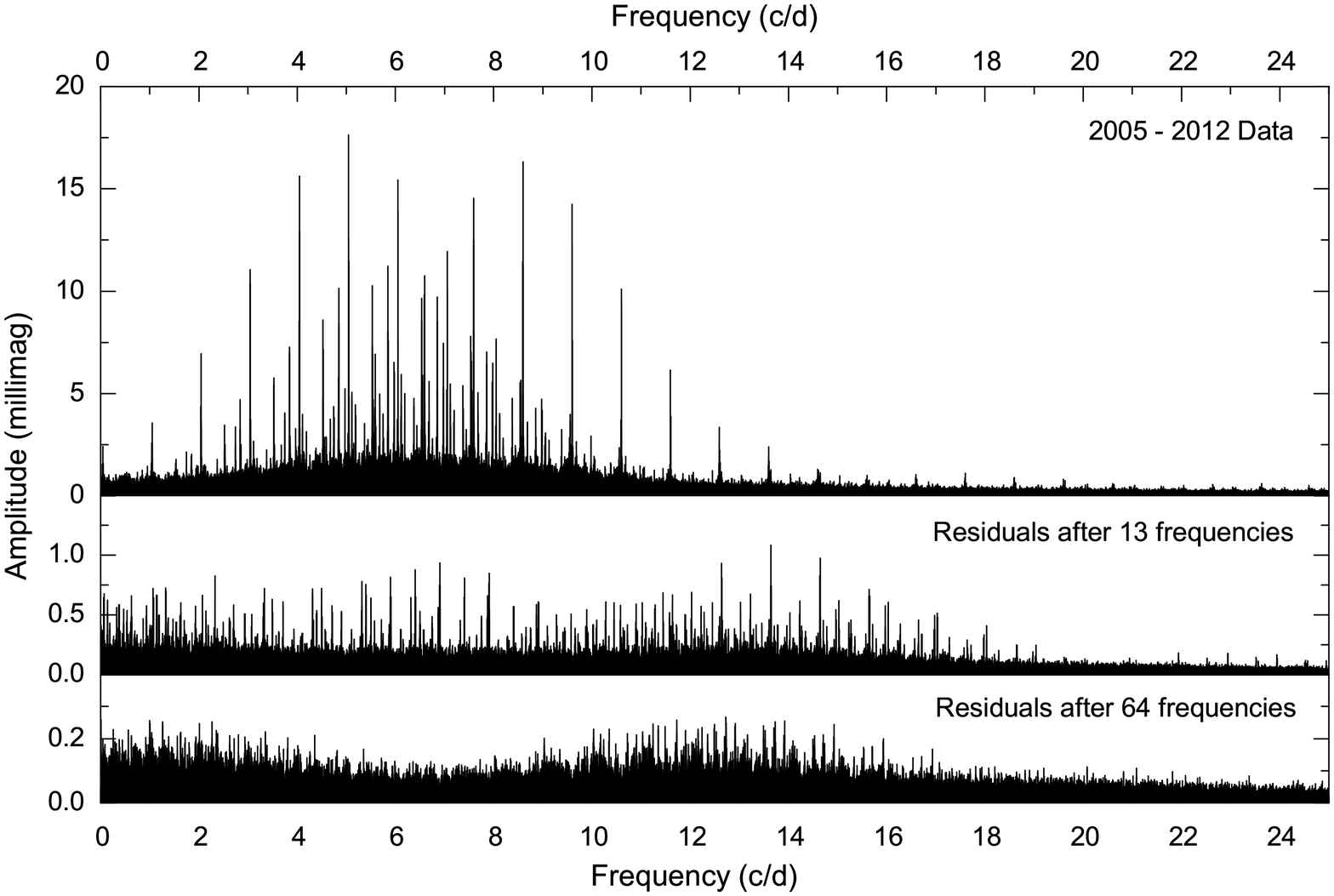}
\caption{Fourier spectra of the 2005-2012 data before and after prewhitening multifrequency solutions. 1\,cycle~d$^{-1}$ aliasing is present because all 702 nights of data were obtained at a single observatory. Different amplitude scales are adopted for the different panels. All 64 detected frequencies are statistically significant. The main p-mode frequency region ranges from 4 to 10 cycle~d$^{-1}$. Most detected peaks in the 0 - 4 and 10 - 20 cycle~d$^{-1}$ region are combinations, $f_i$ $\pm$ $f_j$, of the main pulsation modes $f_i$ and $f_j$.  Additional, statistically less significant, combination peaks can also explain the higher noise seen in those two frequency regions after 64 detected frequencies are prewhitened. The peaks above 20 cycle~d$^{-1}$ are triple combinations.}
\label{Fourier}
\end{figure*}
Between 2005 January 1 and 2012 June 21, more than 700 nights of high-quality photometric data
of 4~CVn were collected with the 0.75m Wolfgang APT telescope located at Washington Camp in Arizona, USA. 
The instrumental and observational details of the telescope and its application to 4~CVn were presented in Breger \& Hiesberger (1999). We measured 4~CVn and two comparison stars
through the Stromgren $y$ and $v$ filters. The comparison stars used were HR~4728 (G9III)
and HR~4843 (F6IV), for which no variability had been detected in a number of previous campaigns.
Furthermore, for many nights the $\gamma$ Dor pulsator HD~108100 was also included as a fourth star in the observing sequence. We found the two dominant frequencies of HD~108100 detected previously by Breger et al. (1997) to be still dominant.

Each cycle of measurements included measurements of all stars in both the $y$ and $v$ passbands and was repeated every three to four minutes.
Because of the large number of comparison-star observations, nightly extinction coefficients could be determined for every filter. They were found to be relatively constant from night
to night. Average seasonal coefficients were adopted (or the nights rejected) for short nights and those nights with variable extinction values. Altogether, data from 702 nights
were used. The dates are shown in Table 1.
 
The times of observations were corrected for the light-time effects in the binary orbit, which has a period of 124.44~d (Schmid et al. 2014). As a check, all pulsation analyses were repeated without light-time corrections. The orbital correction for the primary reduced the residuals of 4~CVn, but the amount of improvement was trivial. The conclusions of this paper, especially concerning the systematic phase shifts of the stellar pulsations, do not depend on whether or not the orbital corrections were applied.

The standard deviation of the difference between the two comparison stars was $\pm$4.1 and $\pm$2.6 millimag in the $v$ and $y$ bandpasses, respectively. This is similar to the residuals determined for the variable star, 4~CVn, after applying the 64-frequency solution determined below ($\pm$3.7 and $\pm$3.0 millimag in $v$ and $y$, respectively). Light curves from the middle of the 2012 campaign are shown in Fig. 1, together with the 64-frequency fits determined below.

\section{Frequency analyses}

For the new 2005--2012 data, the pulsation frequencies were determined using a package of computer
programs with single-frequency and multiple-frequency techniques (PERIOD04;
Lenz \& Breger 2005). The package utilizes both Fourier and multiple-frequency least-squares algorithms, which do not rely on sequential prewhitening.
It is especially suited for extensive data containing up to hundreds of simultaneously excited frequencies.

Owing to the amplitude variability and slow frequency drifts intrinsic to 4~CVn, the frequency analyses were carried out in several stages. These included prewhitening of the
less dominant modes, while permitting amplitude and phase variability of the main pulsation modes. Furthermore, for the modes with amplitudes less than 0.5 millimag, we improved the
signal-to-noise ratios by combining the data from the $v$ and $y$ passbands.
Following Breger et al. (1993), a frequency peak was regarded as statistically significant if it showed an amplitude signal-to-noise ratio of 4.0, which corresponds to a power signal-to-noise ratio $\sim$12.6, or higher. This ad hoc limit of 4.0 is commonly adopted in this type of work and has been examined by extensive numerical simulations (e.g., Kuschnig et al. 1997). However, in some cases, such as for data containing only a single frequency, the limit may be too stringent (Koen 2010).

We identified 64 statistically significant frequencies, which are shown in Table 2. As mentioned earlier, these identifications are independent of whether or not the light-time corrections of the binary orbit are applied. The photometric data also contain a number of additional frequencies near the statistical detection limit. These peaks are mainly situated in the frequency regions in which combination modes are expected. We omitted these statistically less certain peaks, since their detection depends on the details of the chosen noise level and their inclusion would not have added to the discussions in this paper. The Fourier spectra of the 2005--2012 data are illustrated in Fig. 2, which shows the original data, residuals after prewhitening 13 frequencies, and the final residuals after 64 frequencies.

\section{Combination frequencies}

Most $\delta$~Scuti stars show a large number of combination frequencies, $f_i \pm f_j$, with small amplitudes. For the star 4~CVn, the majority of detected frequencies in the previous table are combinations. At least two mechanisms are already known
to create these combinations. The nonlinear response of the medium in the outer layers of the star leads to frequency combinations,
for example, as proposed by Brickhill (1992) and van Kerkwijk et al. (2000) for DA and DB white dwarfs. The observed combination frequencies in the $\delta$~Scuti star 44~Tau were interpreted in this manner (Breger \& Lenz 2008) and the interpretation probably applies to most pulsators. However, resonant excitation of a pulsation mode close to the combination frequency can also lead to a frequency synchronization at exactly $f_i \pm f_j$. In $\delta$~Scuti stars, resonant excitation can lead to amplitudes three orders of magnitude larger than those of the more standard nonlinear case: for the extremely well-studied $\it{KEPLER}$ star KIC~8054146, Breger \& Montgomery (2014) were able to separate the child from parent modes from the phase and amplitude correlations and, thereby, confirm the existence of resonant mode coupling in a $\delta$~Scuti star.

A more complete astrophysical explanation and physical modeling of an observed frequency combination critically
depends on the relative amplitudes of the combination frequency and its parents.
Following our previous analysis for the star 44~Tau, we calculate a combination parameter, $\mu$, which
relates the amplitude of the combination frequency with those of the parent modes,

\begin{equation}
A_{\rm comb} = \mu \cdot A_i \cdot A_j,
\end{equation}
where $A_i$ and $A_j$ are the amplitudes of the parent modes. This definition is similar to the definition of van Kerkwijk et al. (2000). 

We detected 38 combination frequencies in 4~CVn. We  restricted our sample to combinations of the
nine dominant frequencies, $f_1$ to $f_9, $ to minimize
the effects of uncertainties associated with small amplitudes. For these 20 combinations, we calculated the average values of the combination parameters for both the sum and difference frequencies. The results for the $y$ passband, shown in Table 3,  demonstrate the relatively small range in the values of the combination parameter for the selected dominant modes. We also find excellent agreement with the values from a similar analysis of the $\delta$~Scuti star 44~Tau. Consequently, evidence in favor of resonantly coupled modes was not detected in 4 CVn. 

\par
\begin{table}[htb!]
\caption{Relationship between the amplitudes of combination frequencies and parent modes.}
\footnotesize
\begin{tabular}{lccc}
\noalign{\smallskip}
Star    & Combination & Combination parameter, $\mu$ & N\\
& & millimag$^{-1}$\\
\noalign{\smallskip}
\hline
\noalign{\smallskip}
4 CVn & $f_i + f_j$ & 0.0053 $\pm$ 0.0023 & 13\\
& $f_i - f_j$ & 0.0062 $\pm$ 0.0020 & 7\\
44 Tau & $f_i + f_j$ & 0.0046 $\pm$ 0.0025 & 7\\
\noalign{\smallskip}
\hline
\noalign{\smallskip}
\end{tabular}
\newline
The listed uncertainty is the average deviation per single combination frequency.\\
\end{table}

The results presented in the table are by no means complete and are affected by statistical biases; we restricted ourselves to combinations of the dominant modes. Also,
our detection limits discriminate against combinations with extremely small amplitudes, especially at low frequencies (i.e., the  $f_i - f_j$ cases). However, average values can serve
as a rough guide to separate the different physical origins of frequency combination peaks in other $\delta$ Scuti stars.

\begin{figure*}
\includegraphics[bb=10 10 1070 760,width=\textwidth,clip]{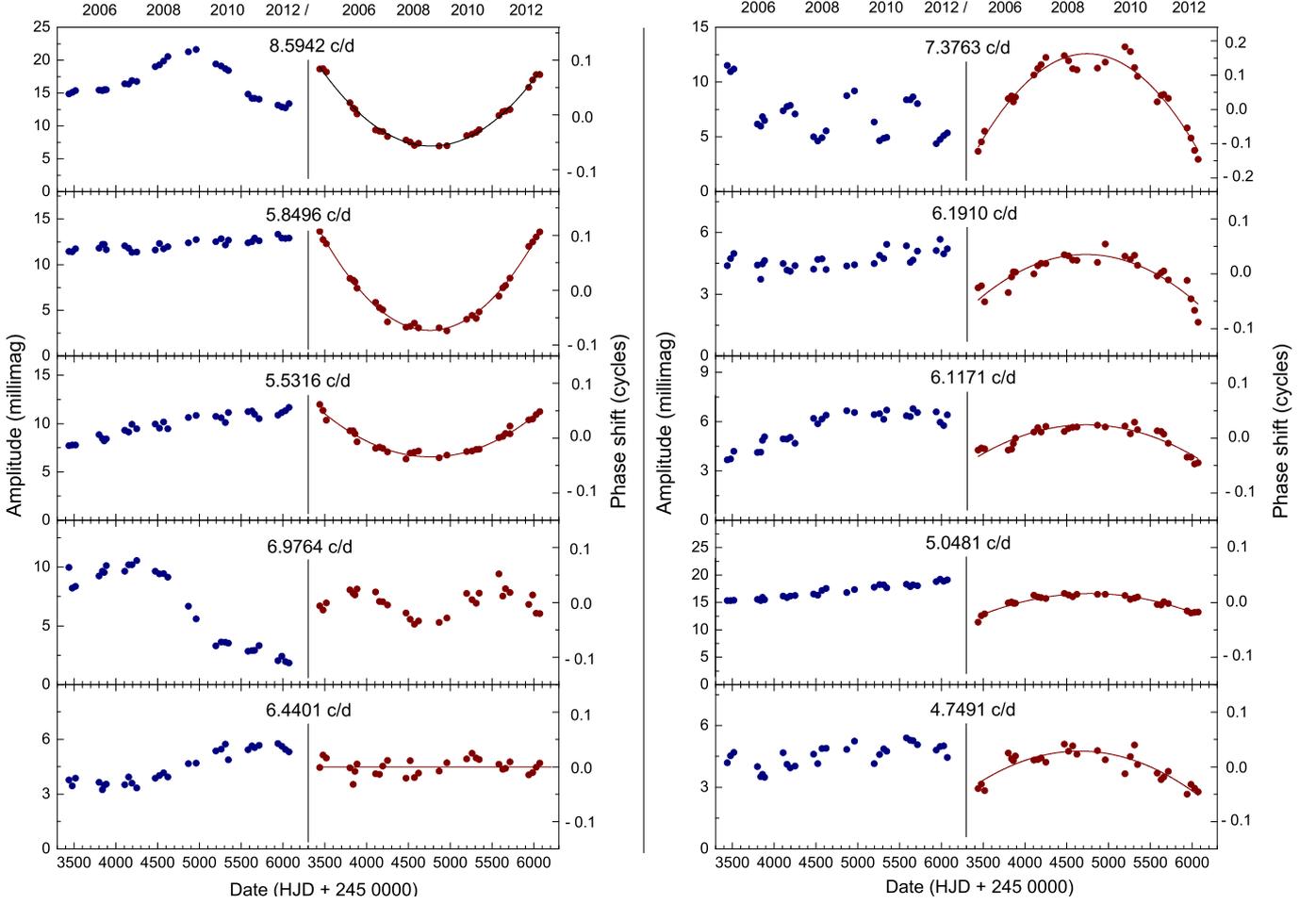}
\caption{(O-C) phase shifts and amplitude variations in the $y$ passband of 10 dominant modes from 2005 to 2012. For each mode, the phase shifts (indicating frequency variations) were determined from an average frequency. The formal uncertainties are generally of the size of the plotted dots and therefore not shown. The solid lines represent least-squares fits of constant frequency changes. The excellent fits for most modes indicate constant frequency (period) changes, suggesting a timescale of at least decades.}
\end{figure*}

\section{Frequency and amplitude variability: 2005 -- 2012}

The standard method to study frequency changes, the so-called (O-C) method, examines frequency changes by assuming a constant frequency and examining the differences in the times between the observations (O) and predictions (C) using the assumed frequency. The method works well only if two conditions are fulfilled: (i) Any lengthy gaps in coverage should not lead to an uncertainty in the number of cycles occurring in the gaps; this condition is fulfilled for the 2005 -- 2012 data.  (ii) The length of time of each subgroup is short enough in order for the frequency to be treated as constant and thereby leading to a correct amplitude value.
This condition is also met when we divide the new data into two to four subgroups for each observing season. Since the amplitude changes and phase shifts, (O-C), are expected to be small and systematic from one subgroup to the next, the method also provides an independent check on the observational errors. To determine the (O-C) shifts, for each pulsation mode we adopted the best frequency for the entire 2005--2012 data and calculated the phase shifts for the large number of subgroups, typically covering between four and six weeks for each subgroup.

 We were able to derive reliable frequency and amplitude changes for 2005 to 2012 for ten dominant modes. These are shown in Fig. 3. The derived phase shifts of two additional modes with average $y$ amplitudes near 6 millimag, $f_7$ and $f_8$, showed large uncertainties and scatter owing to the very small amplitudes over a number of years. These two modes without spectroscopic mode identifications are omitted from further discussion.

\begin{table}
\begin{center}
\caption{Systematic frequency (period) changes, 2005 - 2012}
\begin{tabular}{lcrr}
\hline
\noalign{\smallskip}    
\multicolumn{2}{c}{Frequency, f}  & annual df& (1/P)\,dP/dt \\
& cycle d$^{-1}$ & cycle d$^{-1}$ &yr$^{-1}$\\
\noalign{\smallskip}
\hline
\noalign{\smallskip}
& & x 10$^{-5}$ & x 10$^{-6}$\\
\noalign{\smallskip}
$f_{1}$ &       8.594   &       -5.9    $\pm$   0.1     &       6.9     $\pm$   0.1     \\
$f_{2}$ &       5.048   &       1.7     $\pm$   0.1     &       -3.4    $\pm$   0.2     \\
$f_{3}$ &       5.850   &       -7.6    $\pm$   0.1     &       13.0    $\pm$   0.2     \\
$f_{4}$ &       5.532   &       -3.6    $\pm$   0.1     &       6.5     $\pm$   0.2     \\
$f_{5}$ &       7.376   &       11.7    $\pm$   0.6     &       -15.9   $\pm$   0.8     \\
$f_{6}$ &       6.976   &       -0.5    $\pm$   0.5     &       0.7     $\pm$   0.8     \\
$f_{9}$ &       6.117   &       2.5     $\pm$   0.2     &       -4.1    $\pm$   0.4     \\
$f_{10}$        &       6.191   &       3.7     $\pm$   0.4     &       -5.9    $\pm$   0.6     \\
$f_{11}$        &       6.440   &       -0.1    $\pm$   0.3     &       0.2     $\pm$   0.5     \\
$f_{12}$        &       4.749   &       2.9     $\pm$   0.3     &       -6.2    $\pm$   0.7     \\
\noalign{\smallskip}
\hline
\end{tabular}
\end{center}
\end{table}

\subsection{Regular long-term frequency changes}

The most dominant feature of the observed frequency and amplitude changes is the regular phase shifts seen in eight out of the ten studied frequencies shown in Fig. 3. We also plotted least-squares fits of linearly changing periods for each pulsation mode. The observed changes are smooth with little scatter, confirming that the available data can be used to study the small period changes.

The data can be fit very well with constant frequency changes from 2005 to 1012. The values listed here refer to the $y$ bandpass, but the detected frequency and relative amplitude changes are found to be essentially the same in the $v$ bandpass.

These corresponding values of the period changes are listed in Table 4. Of the eight modes with significant period changes, five modes show period decreases, while three modes have period increases. The values range from (1/P)dP/dt = -16 . 10$^{-6}$ to +13 . 10$^{-6}$ year$^{-1}$ (see Table 4).
The observed period changes are several orders of magnitude larger than those expected from stellar evolution. Since over the eight years the period changes are essentially constant, the timescale of these statistically significant period changes must be on the order of decades or longer.
   
\subsection{Short-term frequency changes of the radial mode}

The radial mode at 6.976 cycle day$^{-1}$ shows small cyclical phase changes in (O-C), although no significant long-term change is found. Our data indicate a timescale of 1690 $\pm$ 50 days. The small variation suggests that the observed period varies on the order of 10$^{-5}$, i.e., it oscillates around its mean value by about 0.3 secs in five years. We can exclude an explanation in terms of light-time effects arising from the 124.44 d  binary orbit because our times were corrected for the light-time effects and no periodicity of 124.44 d were found in the phase-shift data. The origin of this cyclical phase shift is unknown.

\subsection{Amplitude changes}

All dominant modes exhibit amplitude changes over the eight years. There is no common pattern between the amplitude changes of the different pulsation modes. The observed amplitude changes also are not correlated with the period changes and do not depend on the type of pulsation mode.

The radial mode at 6.976 cycle~d$^{-1}$ shows the most dramatic continuous amplitude change in the $y$ passband from 10 millimag in 2007 to 2 millimag in 2012. The decrease was also observed in the spectroscopic amplitude (Schmid et al. 2014).

The amplitude of the retrograde mode at 7.376 cycle~d$^{-1}$ varied with a two-year timescale (P $\sim$ 730 $\pm$ 40 days). A possible explanation of such a regular change is the beating of two close frequencies. For this frequency, a beating model had already been tested to explain the half-cycle phase jump around 1976/7 (Breger 2000). The results at that time were inconclusive.
We tested the beating explanation again by calculating two-frequency models for the new data. The beating models could not fit both the observed amplitude and frequency changes. We conclude that the reason for this amplitude change of the mode at 7.376 cycle~d$^{-1}$ remains unknown.

\section{Discussion}

In the present study, we have presented and analyzed extensive high-precision photometry covering eight years and extracted 64 pulsation frequencies. For a number of pulsation modes constant period changes were found. For six pulsation modes studied in this investigation, the pulsational $m$ values are known (Schmid et al. 2014).
For these six modes, we find a remarkable agreement with the direction of the period change: {\em{the three prograde modes show period increases, while the two known retrograde modes show period decreases}}. Furthermore, the (axisymmetric) radial mode at 6.976 cycle day$^{-1}$ shows no long-term period change, even when data from previous observing campaigns were included.

We conclude that for these modes the sign of the period change is dependent on the azimuthal number of the pulsation modes. Consider a rotationally split $\ell= 1$ triplet. To the first order, the observed frequencies
of the ($\ell = 1, m)$ triplet are given by

\begin{equation}
f_m = f_0 + m \Omega (1 - C),
\end{equation}
where $\Omega$ is the rotational frequency and C is the Ledoux Constant.

In the equation, we can eliminate the axisymmetric mode, $f_0$,  by examining the frequency difference, $\Delta f$, between the $(\ell, m)$ = (1, 1) and (1, -1) modes

\begin{equation}
\Delta f = 2 \Omega (1 - C),
\end{equation}
where any second-order effects have cancelled out due to the subtraction.

The observed dependence on the azimuthal number, $m$, on the sign of the period change leads to a decrease of the rotational separation, $\Delta f$. This, in turn, suggests a decrease in the rotational frequency, $\Omega$. This important discovery will be examined in more detail in a later paper, which will also include all the available photometry covering 40 years.

\begin{acknowledgements}
 
It is a pleasure to thank Michael H. Montgomery for helpful discussions. This investigation has been supported by the Austrian Fonds zur F\"orderung der wissenschaftlichen Forschung
through project P21830-N16.

\end{acknowledgements}

\end{document}